\documentclass[journal]{IEEEtran}

\ifCLASSINFOpdf
  \usepackage[pdftex]{graphicx}
  \DeclareGraphicsExtensions{.pdf,.jpeg,.png}
\else
\fi

\ifCLASSOPTIONcompsoc
  \usepackage[caption=false,font=normalsize,labelfont=sf,textfont=sf]{subfig}
\else
  \usepackage[caption=false,font=footnotesize]{subfig}
\fi

\usepackage{stfloats}
\usepackage{upgreek}
\usepackage{url}
\usepackage{xcolor}
\usepackage{float}

\begin{document}

\title{ Single pixel performance of a 32$\times$32 Ti/Au TES array with broadband X-ray spectra}

\author{Matteo~D'Andrea, Emanuele~Taralli, Hiroki~Akamatsu, Luciano~Gottardi, Kenichiro~Nagayoshi, Kevin~Ravensberg, Marcel~L.~Ridder, Davide~Vaccaro, Cor~P.~de Vries, Martin~de~Wit, Marcel~P.~Bruijn, Ruud~W.~M.~Hoogeveen and Jian-Rong~Gao% <-this % stops a space ~
\thanks{M. D'Andrea is with INAF/IAPS Roma, Via del Fosso del Cavaliere 100, 00133 Rome, Italy. During this research, he was with NWO-I/SRON Utrecht, 3584 CA Utrecht, The Netherlands. e-mail: matteo.dandrea@inaf.it}% <-this % stops a space
\thanks{E. Taralli, H. Akamatsu, L. Gottardi, K. Nagayoshi, K. Ravensberg, M. L. Ridder, D. Vaccaro, C. P. de Vries, M. de Wit, M. P. Bruijn, R. W. M. Hoogeveen and J.-R. Gao are with NWO-I/
SRON Utrecht, 3584 CA Utrecht, The Netherlands.}% <-this % stops a space
\thanks{J.-R. Gao is also with the Faculty of Applied Science, Delft University of Technology, 2600 AA Delft, The Netherlands.}% <-this % stops a space
%\thanks{Manuscript received December 1, 2020; revised December 1, 2020}
}


\maketitle

\begin{abstract}

We are developing a kilo-pixels Ti/Au TES array as a backup option for Athena X-IFU. Here we report on {single-pixel performance} of a 32$\times$32 array operated in a Frequency Division Multiplexing (FDM) readout system, with bias frequencies in the range 1-5 MHz.
{ We have tested the pixels response at several photon energies, by means of a $^{55}$Fe radioactive source (emitting Mn-K$\upalpha$ at 5.9 keV) and a Modulated X-ray Source (MXS, providing Cr-K$\upalpha$ at 5.4 keV and Cu-K$\upalpha$ at 8.0 keV). First, we report the procedure used to perform the detector energy scale calibration, usually achieving a calibration accuracy better than $\sim$ 0.5 eV in the 5.4 - 8.9 keV energy range.} Then, we present the measured energy resolution at the different energies (best single pixel performance: $\Delta$\textit{E}$_\mathrm{FWHM}$ = 2.40 $\pm$ 0.09  eV @ 5.4 keV;  2.53 $\pm$ 0.10  eV @ 5.9 keV; 2.78 $\pm$ 0.16  eV @ 8.0 keV), investigating also the performance dependency from the pixel bias frequency and the count rate. Thanks to long background measurements ($\sim$ 1 day), we finally detected also the Al-K$\upalpha$ line at 1.5 keV, {generated by fluorescence inside the experimental setup}.  We analyzed this line to obtain a first assessment of the {single-pixel} performance also at low energy ($\Delta$\textit{E}$_\mathrm{FWHM}$ = 1.91 eV $\pm$ 0.21 eV @ 1.5 keV), and to evaluate the linearity of the detector response in a large energy band (1.5 - 8.9 keV).

\end{abstract}


\IEEEpeerreviewmaketitle

\section{Introduction}

\IEEEPARstart{A}{thena} is a large X-ray observatory (0.2-12 keV energy band), planned to be launched by ESA in the early 2030s. One of the two focal plane instruments is the X-ray Integral Field Unit (X-IFU) \cite{xifu}, which is a cryogenic spectrometer able to perform simultaneously detailed imaging (5'' angular resolution over a Field of View of 5 arcmin diameter) and high resolution spectroscopy ( $\Delta$\textit{E}$_\mathrm{FWHM}$ $<$ 2.5 eV at E $<$ 7 keV and  $\Delta$\textit{E}$_\mathrm{FWHM}$ $<$ 5 eV at E = 10 keV). The X-IFU will address the \textit{Hot and Energetic Universe} science theme, providing in particular: integral field spectroscopic mapping of the hot plasma in galaxy clusters; weak spectroscopic detection of absorption and emission lines from the Warm Hot Intergalactic Medium (WHIM); physical characterization of the most energetic phenomena in the Universe (e.g. AGN winds and outflows) \cite{xifuscience}. The core of the instrument is a large array of $\sim$ 3000 Mo/Au Transition Edge Sensor (TES) microcalorimeters, provided by NASA-Goddard and operated at a 50 mK thermal bath temperature {\cite{nasa1}\cite{nasa2}}. 

We are developing a Ti/Au TES array as {a} European backup technology for Athena X-IFU. In the recent past, { the work has been focused} in the optimization of the pixels design \cite{sron1}\cite{sron2}\cite{sron3}, and their characterization under both AC and DC bias \cite{sronacdc}. Furthermore, kilo-pixel arrays have been produced and operated in a Frequency Division Multiplexing (FDM) readout system, in order to evaluate the homogeneity of the pixels characteristics and performances \cite{sron4}\cite{taralli}\cite{taralliasc}. { In these tests,} the detector response has been usually evaluated at 5.9 keV, by using the Mn K$\upalpha$ calibration line generated by means of a radioactive $^{55}$Fe source. 

{{ In this work, we present single-pixel performance at different photon energies (i.e. 1.5 keV, 5.4 keV, 5.9 keV and 8.9 keV), and we preliminary investigate the achievable energy scale calibration accuracy, the detector linearity and the count-rate effect on the performance.}

\section{Experimental Setup}

The detector under test is a uniform 32$\times$32 pixels array, with 240$\times$240 $\upmu$m$^2$ Au absorbers (2.35 $\upmu$m thickness). The TESs have an aspect ratio (lenght $\times$ width) of 140$\times$30 $\upmu$m$^2$, and consist of Ti/Au (35 nm/ 200 nm) bilayers with a critical temperature \textit{T}$_\mathrm{C}$ $\sim$ 90 mK and a normal resistance \textit{R}$_\mathrm{N}$ $\sim$120 m$\Omega$. The pixel heat capacity is C $\sim$ 0.85 pJ/K at \textit{T}$_\mathrm{C}$, and the expected thermal conductance is G $\sim$ 95 pW/K at \textit{T}$_\mathrm{C}$. More details about the detector design and fabrication can be found in \cite{sron4}, while a deep characterization of the array has been reported in \cite{taralli} and \cite{taralliasc}.

The detector has been operated in our FDM readout system \cite{fdm1} \cite{fdm2}. Under the FDM scheme, TES are coupled to high-Q LC filters \cite{lcfilters} and AC biased at MHz frequencies. The TES signals are then summed and amplified by a low-noise two-stage SQUID {amplifier chain}, and finally demodulated at room temperature by digital electronics. The main parts of the cold stage setup are shown in Figure \ref{setup}. We have used a 18 channels LC-filter chip, with a coil inductance L = 2 $\upmu$H and a DC connector chip. The pixels bias frequencies are between 1 and 5 MHz. The {SQUID} amplifiers are a VTT J3 (Front End SQUID) and a VTT F5 (Amp SQUID). 

\begin{figure}[!ht]
\centering
\includegraphics[width=3.in]{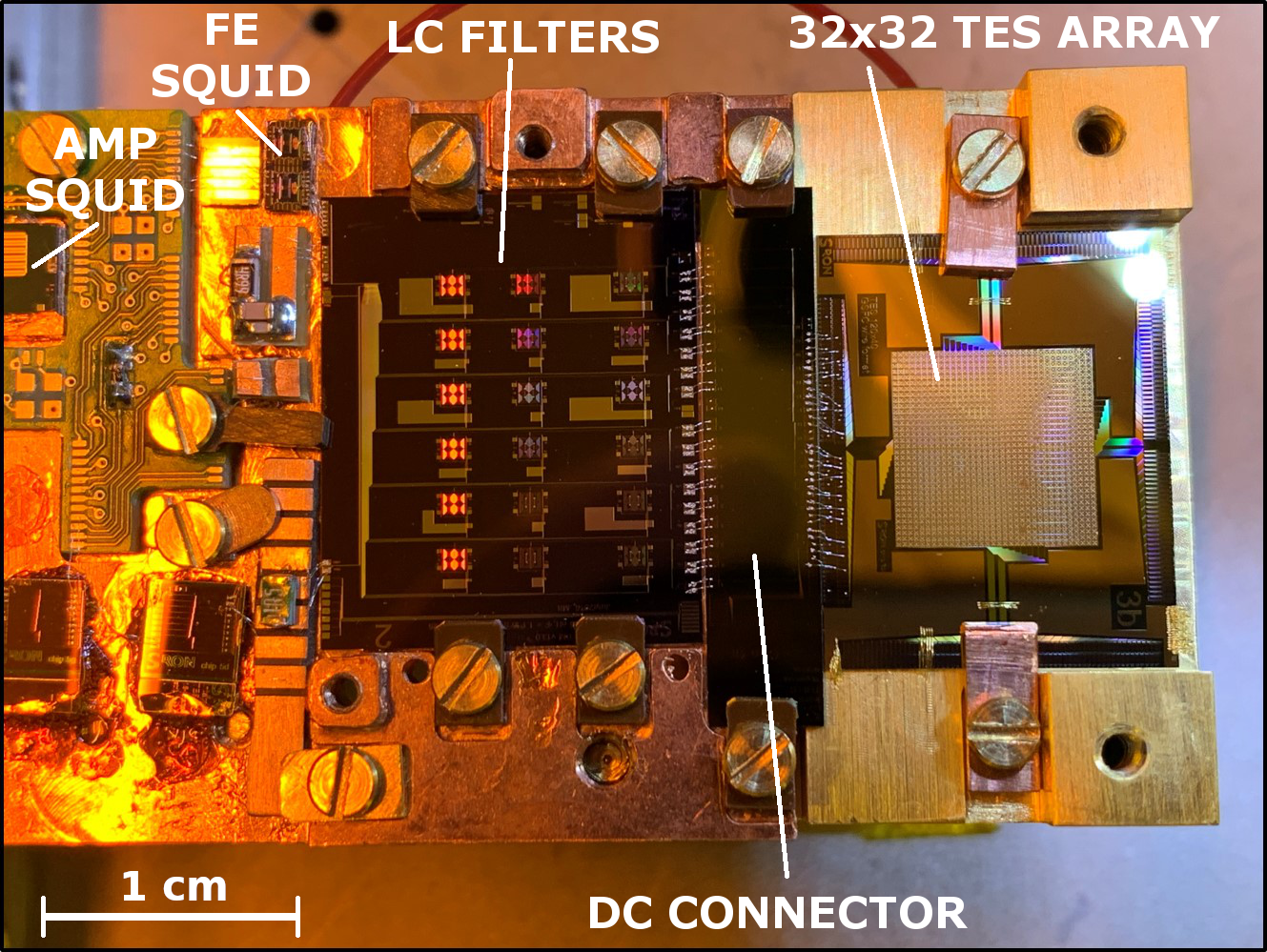}
\caption{Picture of the setup (cold stage) used to operate the kilo-pixel array. The labels indicate the main components.}
\label{setup}
\end{figure}

The setup has been installed in a dilution refrigerator, and thermoregulated at a temperature of 40 mK. It has been suspended with respect to the cryostat mixing chamber via kevlar strings, in order to avoid microvibrations due to the pulse tube operation, and magnetically shielded by a superconducting aluminum shield (at cold), and by a mu-metal shield (outside the cryostat).

The detector has been illuminated both by a radioactive $^{55}$Fe source (from the back side) and a {Modulated X-ray Source} (MXS, from the front side). The MXS is a device able to produce X-rays with an adjustable count rate, by hitting target materials with accelerated electron, which are emitted by a photocathode stimulated by UV leds. Details about the MXS design are reported in \cite{mxs}. The MXS has been vacuum tight connected to an aperture in the cryostat Outer Vacuum Chamber. All the thermal shields inside the cryostat have been provided with apertures covered with 20 microns thick mylar coated with 200 nm Al. The main purpose of such mylar windows is to block the stray light and let the X-ray pass through the cryostat.

\begin{figure}[!b]
\centering
\includegraphics[width=3.in]{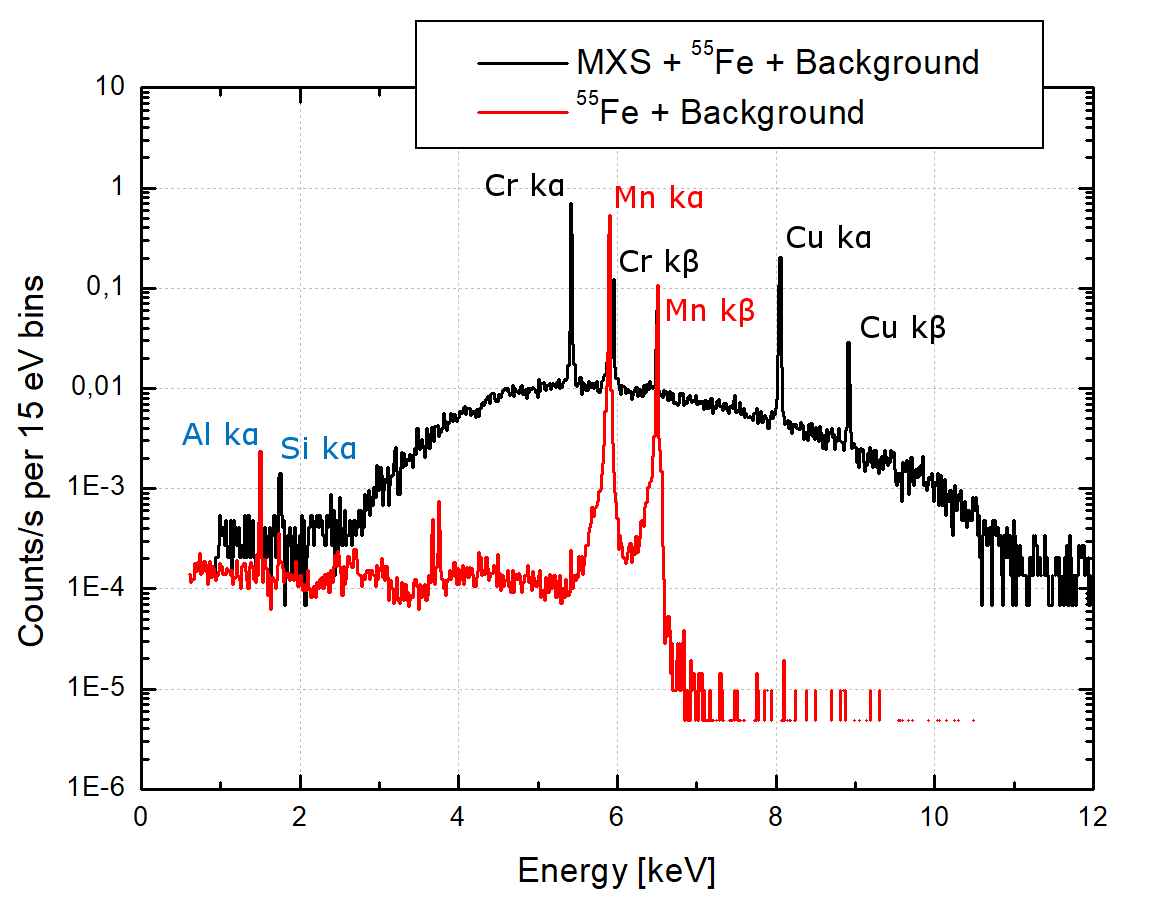}
\caption{Energy spectrum measured by Px8 (2.7 MHz bias frequency) with the MXS turned ON (black curve) and OFF (red curve). }
\label{mxs}
\end{figure}

Figure \ref{mxs} shows the comparison of the energy spectra acquired by one of the detector pixels (Px8, 2.7 MHz bias frequency) with the MXS turned ON and OFF. The Mn-K$\upalpha$ (5.9 keV) and  Mn-K$\upbeta$ (6.4 keV) lines are emitted by the $^{55}$Fe source. The MXS provides the Cr-K$\upalpha$ (5.4 keV), Cr-K$\upbeta$ (5.9 keV), Cuk-K$\upalpha$ (8.0 keV) and Cu-K$\upbeta$ (8.9 keV) lines, and the bremsstrahlung continuum. In addition, two low-energy lines are slightly visible in the spectra, generated by fluorescence inside the setup: the Al-K$\upalpha$ at 1.5 keV (generated in the aluminum foils used to attenuate the $^{55}$Fe source), and the Si-K$\upalpha$ at 1.7 keV (generated in the detector substrate).

\section{Analysis procedure \& Energy scale calibration}

We have operated the TES array in single-pixel mode configuration, biasing  one pixel at a time, while all the others are left in the superconducting state. We have configured the MXS in order to have a count-rate of $\sim$ 3 counts per second per pixel, and collected 50000 pulses per measure. To analyze the data, we have then proceeded as follows:

\begin{itemize}
\item[1 -] We have generated pulse and noise templates, and performed the pulses optimal filtering. The pulse template has been generated by averaging 100 pulses in the 5.9 keV line;
\item[2 -] We have linearly calibrated the energy scale by roughly using the most prominent line in the spectrum (i.e. the Mn-K$\upalpha$ at 5.9 keV);
\item[3 -] We have preliminarily fitted all the main K$\upalpha$ and K$\upbeta$ lines in the spectrum (Figure~\ref{prem_fit}). The intrinsic line shapes have been described according to the ASTRO-H SXS Calibration Database for Line Fit \cite{hitomi};

\begin{figure}[!h]
\centering
\includegraphics[width=3.4in]{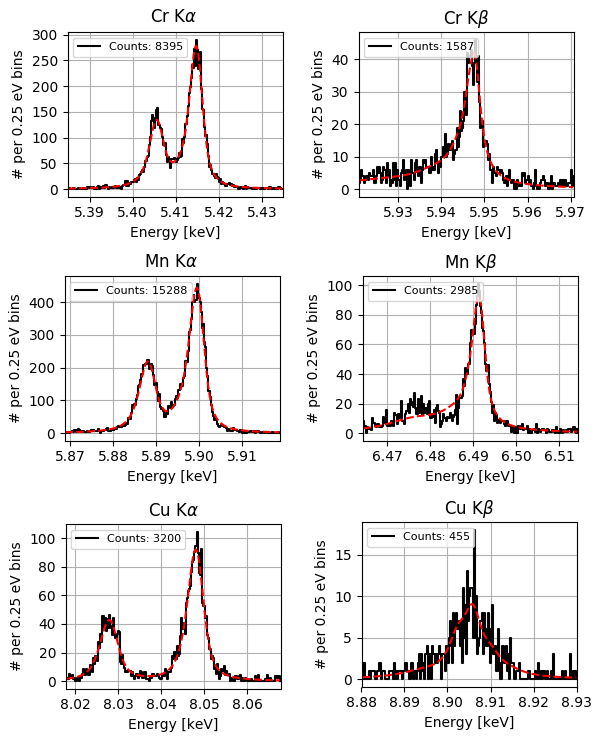}
\caption{Preliminary fit of the main K$\upalpha$ and K$\upbeta$ lines (generated by MXS and internal $^{55}$Fe X-ray source) acquired by Px1 biased at 1.1 MHz.}
\label{prem_fit}
\end{figure}

\setcounter{figure}{4}
\begin{figure*}[!b]
\centering
\includegraphics[width=2.3in]{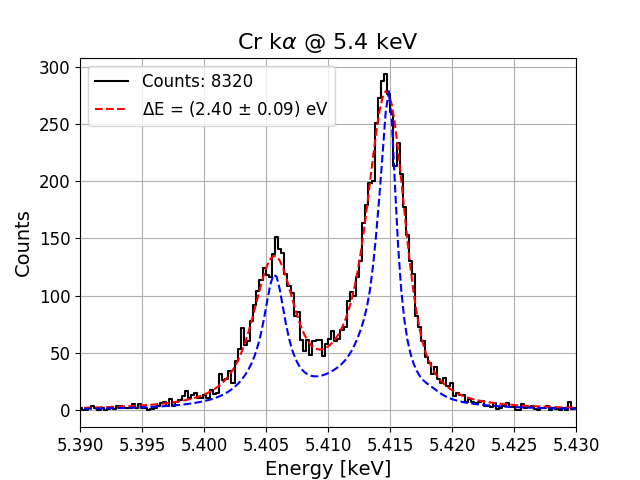}%
\includegraphics[width=2.3in]{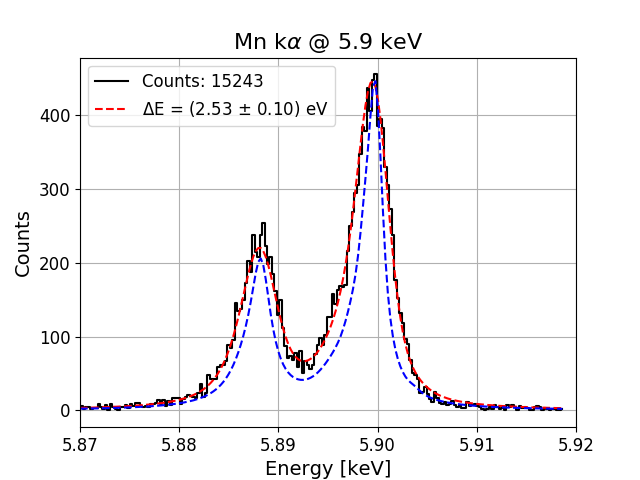}%
\includegraphics[width=2.3in]{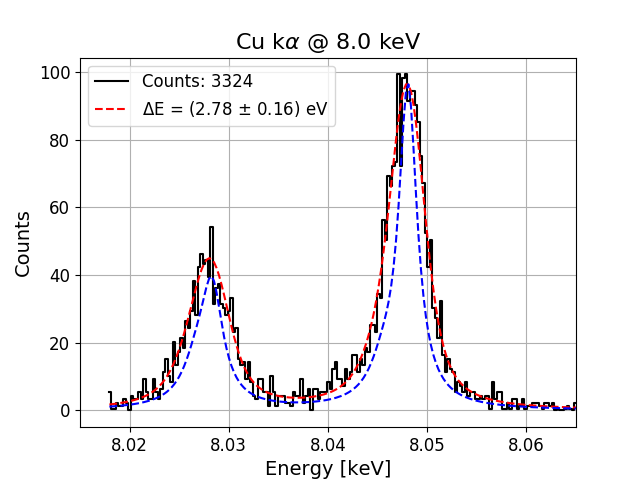}%
\caption{Best fit to data (red lines) for each detected K$\upalpha$ line (MXS and internal 55Fe X-ray source) for Px1 biased at @ 1.1 MHz. The lines intrinsic shape (blue lines) has been described according to \cite{hitomi}.}
\label{bestfits}
\end{figure*}

\setcounter{figure}{3}
\begin{figure}[H]
\centering
\includegraphics[width=3.4in]{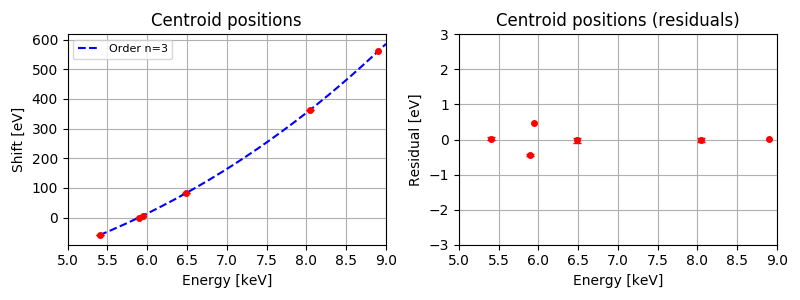}
\caption{(Left) Measured energy shift of the detected lines as a function of the energy, as obtained by the preliminary fit (data from Px1 biased at 1.1 MHz). The blue dashed lines is a 3th order polynomial fit of the data. (Right) Residuals of the 3th order polynomial fit. The fit accuracy is better than 0.5 eV. }
\label{calibration}
\end{figure}

\item[4 -] From the preliminary fits, we obtained the energy shift of each line with respect to its real position. We have then fitted the measured shifts as a function of the energy with a 3th order polynomial (Figure~\ref{calibration} - Left), and used the obtained Shift (E) function to correct the energy scale calibration. Note that the fit residuals (Figure~\ref{calibration} - Right) represent the final energy calibration accuracy. With this method, we have usually achieved an accuracy better than {$\sim$} 0.5 eV in the energy range between 5.4 and 8.9 keV (for comparison, the X-IFU will require a Mean Knowledge Error of the energy scale smaller than 0.4 eV in the 0.2 to 7 keV range \cite{casA});

\item[5 -] We have finally re-fitted the K$\upalpha$ lines in the spectrum to assess the detector energy resolution at the different energies. Before this final fit procedure, we performed a simple bremsstrahlung subtraction from the lines ($\sim$1 count per bin subtracted). 

\end{itemize}

\section{Energy resolution}
The best measured energy resolutions at the different energies are shown in Figure~\ref{bestfits}. The spectra have been acquired using the Px1 (1.1 MHz bias frequency) {in single-pixel readout}, and they show $\Delta$\textit{E}$_\mathrm{FWHM}$ = (2.40 $\pm$ 0.09)  eV @ 5.4 keV;  (2.53 $\pm$ 0.10)  eV @ 5.9 keV and (2.78 $\pm$ 0.16)  eV @ 8.0 keV.  The performance achieved at 5.9 keV is consistent with the one previously reported for this array \cite{taralli}, validating both the data analysis procedure and the new experimental setup with the MXS.

We have characterized several pixels during this measurement run, obtaining consistent results. Figure~\ref{freqperformance} shows the energy resolution measured for Cr-K$\upalpha$, Mn-K$\upalpha$ and Cu-K$\upalpha$ comparing 3 pixels at different bias frequencies: Px1 at 1.1 MHz, Px8 at 2.7 MHz and Px16 at 4.6 MHz (i.e. low, mid and high bias frequency). {All the measurements have been performed in single-pixel readout}. The dashed black line represents the final X-IFU requirement at instrument level, and it is overplotted just for reference.
Note that, {for each pixel}, the energy resolution degradation between 5.9 and 8.0 keV ($< 0.3$ eV) is lower than the one requested by the X-IFU requirements ($\sim 0.8$ eV).

\setcounter{figure}{5}
\begin{figure}[h]
\centering
\includegraphics[width=3.in]{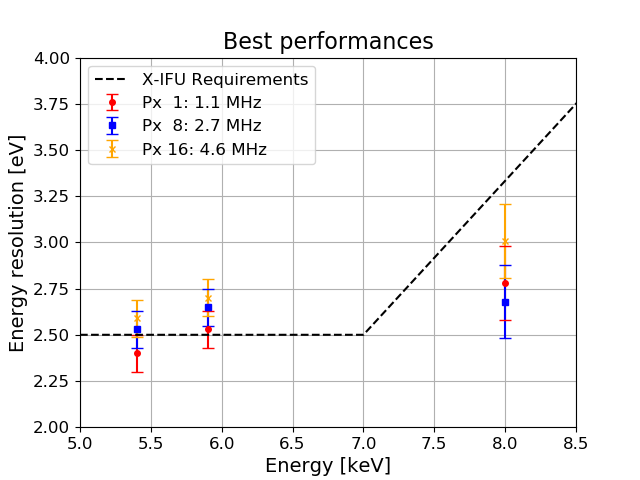}
\caption{Energy resolution as a function of the MXS emission line energies for 3 pixels at different bias frequencies: Px1 at 1.1 MHz, Px8 at 2.7 MHz and Px16 at 4.6 MHz. The dashed black line represents the X-IFU requirement at instrument level.}
\label{freqperformance}
\end{figure}

\section{Low-energy response \& linearity}

To investigate the detector behavior also at low energy, we performed long measurements ( $\sim$ 1 day of observing time) to detect the background Al and Si fluorescence lines generated inside the setup (slightly visible in the spectra in Figure \ref{mxs}). This allowed us to estimate the energy resolution at 1.5 keV, and to perform a first assessment of the detector linearity in a large energy band.

Figure \ref{fital} shows the Al-K$\upalpha$ complex acquired in a 20 hour acquisition with Px 8 (2.7 MHz bias frequency). The measured energy resolution is $\Delta$\textit{E}$_\mathrm{FWHM}$ = (1.91 $\pm$ 0.21)  eV @ 1.5 keV. In this case, the proper energy scale calibration has been obtained by linearly fitting the centroid position of the Al-K$\upalpha$ (1.49 keV) and the Si-K$\upalpha$ (1.74 keV) lines. Because of the low Si-K$\upalpha$ count rate ($\sim$ 80 counts in 20 hours), it has been not possible to assess the detector performance also at that energy. 

\begin{figure}[!t]
\centering
\includegraphics[width=3.in]{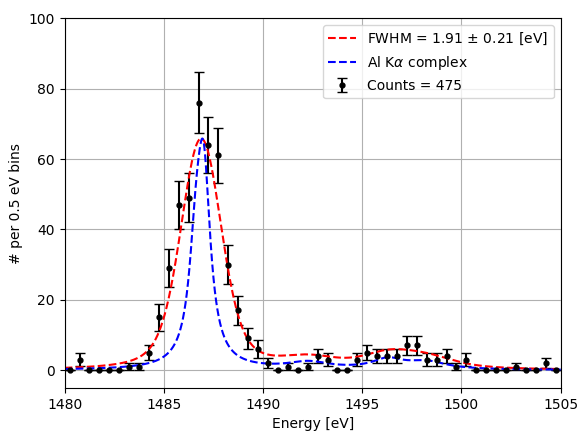}
\caption{Best fit (red line) of the detected Al K$\upalpha$ line for Px8 biased at @ 2.7 MHz. The line intrinsic shape (blue line) has been described according to \cite{hitomi}.}
\label{fital}
\end{figure}

Considering the whole energy range we have explored so far (from 1.5 to 8.9 keV), we can estimate the linearity of the detector. Figure \ref{linearity} shows the optimal filtering output parameter (i.e. the normalized area of the optimal filtered pulses, which we use as optimal energy estimator) obtained for each detected emission lines as a function of the real lines energy. The dashed blue line is the linear trend calibrated on the detected low energy lines. We observe a deviation from the linearity of 8\% at 5.9 keV and 15\% at 8.9 keV.

\begin{figure}[!h]
\centering
\includegraphics[width=3.in]{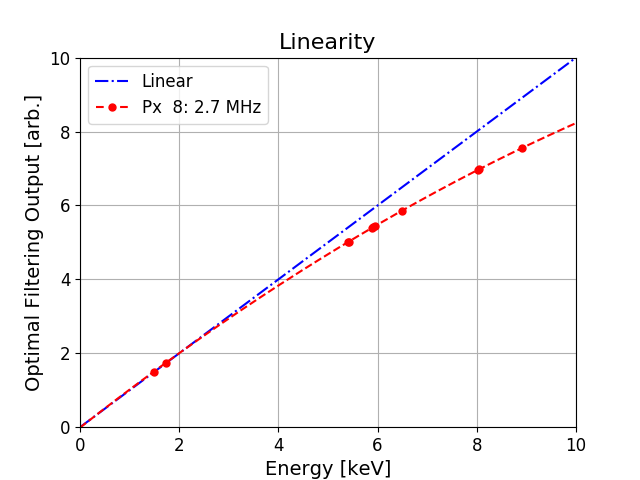}
\caption{Detector linearity in the range from 1.5 to 9 keV. The optimal filtering output is shown as a function of the detected line energies (red points). The dashed blue line is the linear trend calibrated on the low energy lines.}
\label{linearity}
\end{figure}

\section{Count rate effect}

Finally, we have performed a preliminary investigation of the energy resolution trend as a function of the count rate on the {pixels}. The emission rate of the MXS can be indeed increased by increasing the current in the UV-LEDs. Figure~\ref{countrate} shows the measured energy resolution of the Cr-K$\upalpha$ line (5.4 keV) as a function of the count rate. The red points are averaged values over the three pixels biased at 1.1, 2.7 and 4.6 MHz, {and operated in single-pixel readout}. The blue line represent the fraction of good events used to fit the spectra. We performed a very simple pulse selection, in which we reject events where multiple pulses are detected within our optimal record lenght (26 ms). We observe that for up to around 10 counts per second no significant performance degradation is visible, and that the fraction of high resolution events is higher than 80\%. {Just} for reference, consider that one of the X-IFU requirements is to observe the average flux of the supernova remnant CasA ($\sim$ 2.1 cps/pixel) with undegraded energy resolution and $>$ 80\% high resolution events \cite{casA}. 

{Note that here, in single-pixel readout, the primary mechanism of resolution degradation is the induced thermal instability on the pixel baseline. To proper evaluate the effect of the thermal crosstalk due to the neighbor pixels, we have performed also a dedicated measurements in multiplexed readout. This activity is reported in another paper in this issue \cite{taralliasc}.}

\begin{figure}[!t]
\centering
\includegraphics[width=3.in]{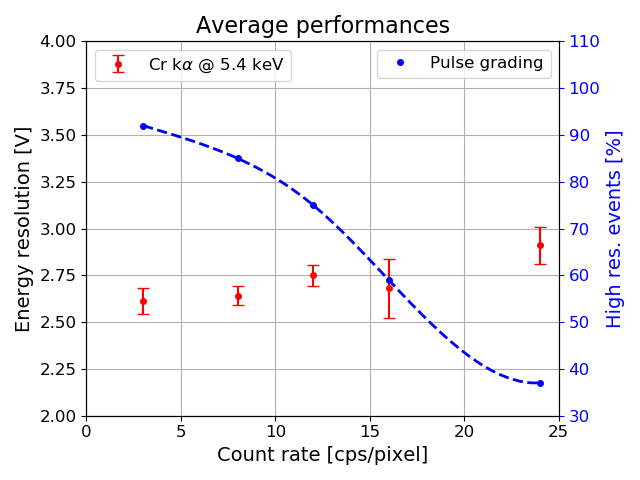}
\caption{Energy resolution of the Cr-K$\upalpha$ line (5.4 keV) as a function of the count rate. Red points are average values over the three pixels biased at 1.1, 2.7 and 4.6 MHz. The blue points indicate the fraction of good events used to fit the spectra. Consider that on the X-IFU 30 cps/pixel roughly correspond to 1 Crab flux.}
\label{countrate}
\end{figure}

\section{Conclusion}

We are developing a large Ti/Au TES array as a backup technology for ATHENA X-IFU. In this work, we have presented { the single-pixel performance at different X-ray energies} of a uniform 32x32 pixels prototype, which has been operated in the FDM readout system. The achieved energy scale calibration accuracy is {usually} better than {$\sim$} 0.5 eV in the range 5.4 - 8.9 keV. We have measured as best single pixel performance: $\Delta$\textit{E}$_\mathrm{FWHM}$ = (1.91 $\pm$ 0.21)  eV @ 1.5 keV, (2.40 $\pm$ 0.09)  eV @ 5.4 keV,  (2.53 $\pm$ 0.10)  eV @ 5.9 keV, (2.78 $\pm$ 0.16) eV @ 8.0 keV. Finally, we have not noticed performance degradation as a function of the count rate, up to $\sim$ 10 count/s. 

These results represent a promising first reference for the kilo-pixel array behavior in a large energy band, as well as conclude an extensive array characterization including performance and characteristics uniformity \cite{taralli} and thermal crosstalk minimization \cite{taralliasc}. A further optimization of the pixel design, already showing significant performance improvement \cite{sron3}, could lead in the near future to demonstrate a full array inside the X-IFU requirements.

\section*{Acknowledgment}

This work is partly funded by European Space Agency (ESA) and coordinated with other European efforts under ESA CTP contract ITT AO/1-7947/14/NL/BW. It has also received funding from the European Union’s Horizon 2020 Programme under the AHEAD (Activities for the High-Energy Astrophysics Domain) project with grant agreement number 654215.

\ifCLASSOPTIONcaptionsoff
  \newpage
\fi



%\bibliographystyle{IEEEtran}

%\newpage

\end{document}